\documentclass{aastex}

\usepackage{graphicx}

\usepackage{emulateapj5}
\usepackage{times}

\makeatletter
\newenvironment{inlinetable}{%
\def\@captype{table}%
\noindent\begin{minipage}{0.999\linewidth}\begin{center}\footnotesize}
{\end{center}\end{minipage}\smallskip}

\newenvironment{inlinefigure}{%
\def\@captype{figure}%
\noindent\begin{minipage}{0.999\linewidth}\begin{center}}
{\end{center}\end{minipage}\smallskip}
\makeatother

\usepackage{mathptmx}

\def\keV{ke\kern-0.05emV}

\newcommand{\chandraintitle}{{\itshape\footnotesize CHANDRA}\/}
\newcommand{\ROSAT}{\textsl{ROSAT}}
\newcommand{\NGC}{NGC\,}
\renewcommand{\arcsec}{\ensuremath{''}}
\renewcommand{\mp}{m_{\!p}}

\begin{document}

\submitted{Submitted to ApJ Letters, February 27 2001}

\title{Zooming in on the Coma cluster with \emph{Chandra}: compressed
  warm gas in the brightest cluster galaxies}

\author{A.\ Vikhlinin, 
  M.\ Markevitch,
  W.\ Forman, and C. Jones}
\affil{Harvard-Smithsonian Center for Astrophysics, 60 Garden St.,
Cambridge, MA 02138;\\ avikhlinin@cfa.harvard.edu}


\shorttitle{ZOOMING IN ON COMA WITH \emph{CHANDRA}}
\shortauthors{VIKHLININ ET AL.}

\begin{abstract}
  The \emph{Chandra} image of the central region of the Coma cluster reveals
  that both its dominant galaxies, \NGC4874 and \NGC4889, retain the central
  parts of their X-ray gas coronae. The interstellar gas with a temperature
  of 1--2~\keV\ is confined by the hot intergalactic medium of the Coma
  cluster into compact clouds (only 3~kpc in radius) containing
  $10^8\,M_\odot$ of gas. The physical state of the gas in these clouds
  appears to be determined by a delicate balance between radiative cooling
  and suppressed (by a factor of 30--100) heat conduction through the
  interface between these clouds and the hot cluster gas.
\end{abstract}


\section{Introduction}

\ROSAT\ observations showed that the Coma cluster, once thought to be an
archetype of a relaxed system, contains substructures on a wide range of
linear scales (Briel, Henry \& B\"ohringer~1992, White, Briel \& Henry~1993,
Vikhlinin, Forman \& Jones 1997). The detected structures include
100~kpc-scale X-ray enhancements around the two dominant cluster galaxies,
\NGC4874 and \NGC4889 (Vikhlinin, Forman, \& Jones~1994). Recent XMM
observations confirmed these findings and also revealed unresolved X-ray
sources coincident with these galaxies (Arnaud et al.\ 2001, Briel et al.\ 
2001). In this \emph{Letter}, we resolve these sources with \emph{Chandra}.
Both sources are extended arcsec-scale ($\approx 3$~kpc) remnants of
galactic X-ray coronae with temperatures 1--2~keV, compressed by the hot,
9~keV, gas of the Coma cluster. Their very existence poses interesting
physical questions.

The problem of thermal evaporation of cold clouds embedded in a hot medium
has been considered in a number of papers (Cowie \& McKee 1977, McKee \&
Cowie 1977, Balbus \& McKee 1982, McKee \& Begelman 1990, Bandiera \& Chen
1994 among others). This problem is relevant for the survival of cold clouds
in multi-phase cluster cooling flows (e.g., B\"ohringer \& Fabian 1989,
Fabian, Canizares, \& B\"ohringer 1994). The general consensus emerging from
these studies, as summarized by Fabian (1994), is that heat conduction must be
suppressed by one to two orders of magnitude below the Spitzer value,
otherwise the radiative cooling of the gas will be overcome by heat
conduction.  However, the direct evidence for suppression of heat
conductivity in the intracluster medium (ICM) is scarce (one of a few
examples can be found in Bechtold et al.\ 1983). The \emph{Chandra}
observations of the interstellar medium in the Coma galaxies presented here
allow a direct study of cold clouds embedded in a hot intracluster medium.

We use $H_0=50$~km~s$^{-1}$~kpc$^{-1}$, which implies a linear scale of
0.68~kpc per arcsec at the Coma distance ($d=140$\,Mpc).

\section{\chandraintitle\ observations}

The Coma cluster was observed with ACIS in the fall of 1999 in a series of 6
pointings with individual exposures of $10$~ksec. The pointing position was
kept the same during these observations, with \NGC4874 and \NGC4874 located
at off-axis distances of 2.8\arcmin\ and 4.0\arcmin, respectively. The
detector was moved in the focal plane between the pointings, so that the
galaxies were observed by both ACIS-I (4 times) and ACIS-S (2 times). There
were no background flares in the Coma observations. In one pointing,
\NGC4874 was located in the gap between the ACIS-I chips. The total exposure
was 48~ksec for \NGC4874 and 56~ksec for \NGC4889.

The central part of the composite ACIS-I image is shown in
Fig.~\ref{fig:big}. Notice the faint arcmin-scale X-ray enhancements around
\NGC4874 and 4889, previously detected by \ROSAT. At the center of these
extended structures there are bright (500 counts) compact sources
coincident with the optical positions of the galaxies. The observed
0.5--2~keV flux is $3.9\pm0.1\times10^{-14}$~erg~s$^{-1}$~cm$^{-2}$ for
\NGC4874\ and $3.2\pm0.1\times10^{-14}$~erg~s$^{-1}$~cm$^{-2}$ for \NGC4889,
corresponding to luminosities in the same band of $9.1\times10^{40}$ and
$7.6\times10^{40}$~erg~s$^{-1}$.

The enlarged X-ray and optical images of \NGC4874 and 4889 are shown in
Fig.~\ref{fig:panels}. The X-ray sources associated with the galaxies are
extended, but unlike most ellipticals, the X-ray emission is much more
compact than the distribution of optical light. Below we discuss these
interesting objects in detail.

\section{Spectral analysis}

The spectra of the \NGC4874 and 4889 sources were extracted within a radius
of 7\arcsec\ (which encompasses most of the observed X-ray flux). Since we
are interested in the intrinsic properties of the galaxies, we collected the
background spectra in the 15\arcsec--30\arcsec\ annuli around them, so it
includes the true background as well as the emission of the Coma cluster.
The source, background spectra, and the response files were generated
individually for each pointing. We then co-added the observed spectra and
averaged the response files for each source.

\begin{figure*}[t]
  
  
  
  \centerline{\includegraphics{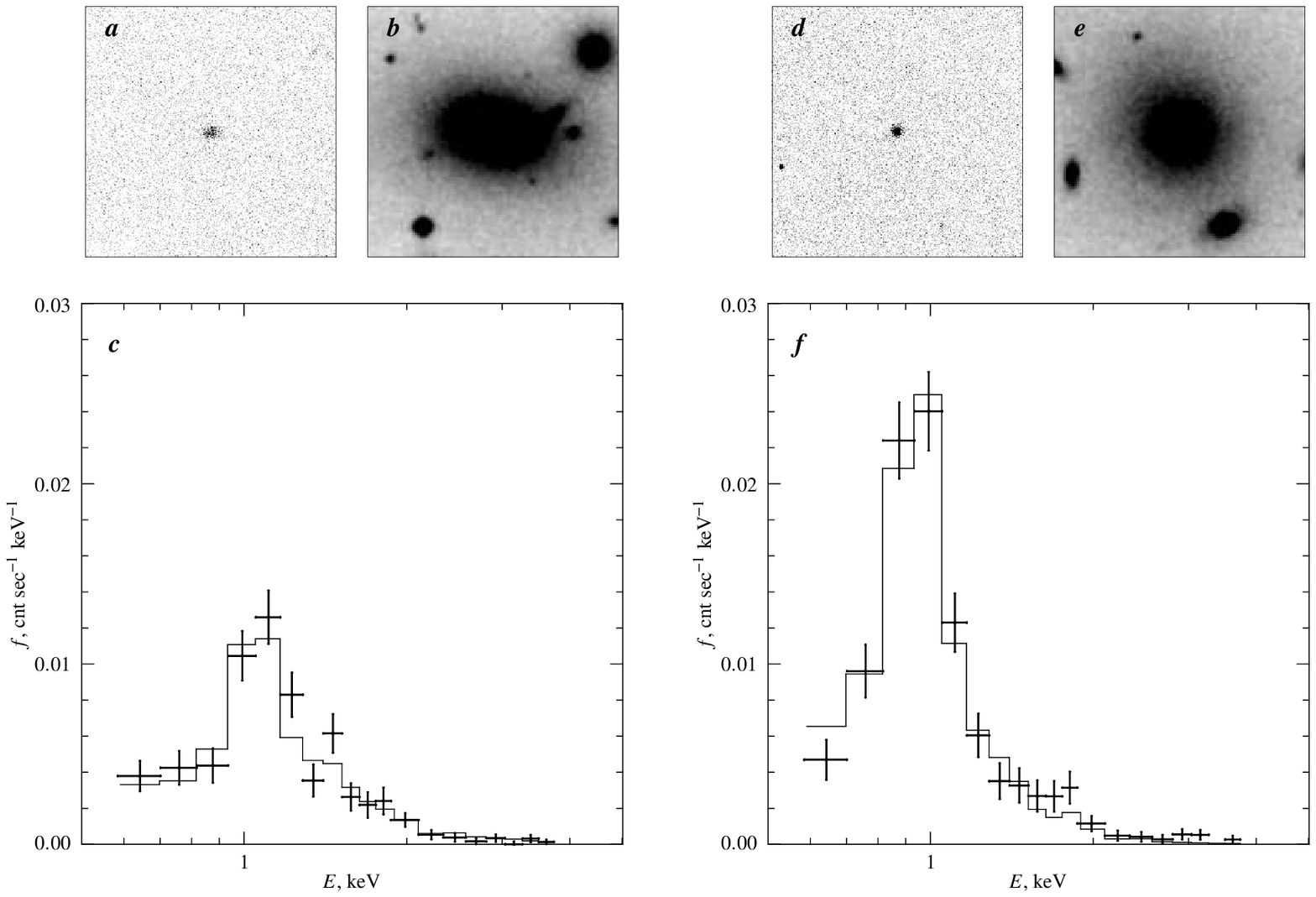}}

  \caption{Images and spectra of \NGC4889 (\emph{a}, \emph{b}, \emph{c}) and
    \NGC4874 (\emph{d}, \emph{e}, \emph{f}). The X-ray (0.5--2~keV) and
    optical images have the same angular size, $2'\times2'$. Solid
    histrograms is panels \emph{c} and \emph{f} show the best-fit model with
    parameters from Table~\ref{tab:spectra}.}
  \label{fig:panels}
  
\end{figure*}

In the spectra for both galaxies (Fig.~\ref{fig:panels}), one immediately
notices strong emission lines near 1~keV --- a key signature of a
1~keV-temperature plasma. Indeed, the spectra are well fit by the
Mewe-Kaastra-Liedahl model in XSPEC with the absorption fixed at the
Galactic value $N_H=9\times10^{19}$~cm$^{-2}$. Therefore, the observed X-ray
emission is most likely produced by warm interstellar gas. The X-ray
emission cannot be due to the integrated emission of stars because its
spatial distribution does not follow the optical light
(Fig.~\ref{fig:panels} and \ref{fig:profile:4874}). Any significant
contribution by low-mass X-ray binaries (LMXB) can be ruled out because it
also should follow the starlight and have a very different (hard) spectrum
(Sarazin, Irwin, \& Bregman 2001).

\begin{inlinefigure}
  \centerline{\includegraphics[width=0.95\linewidth]{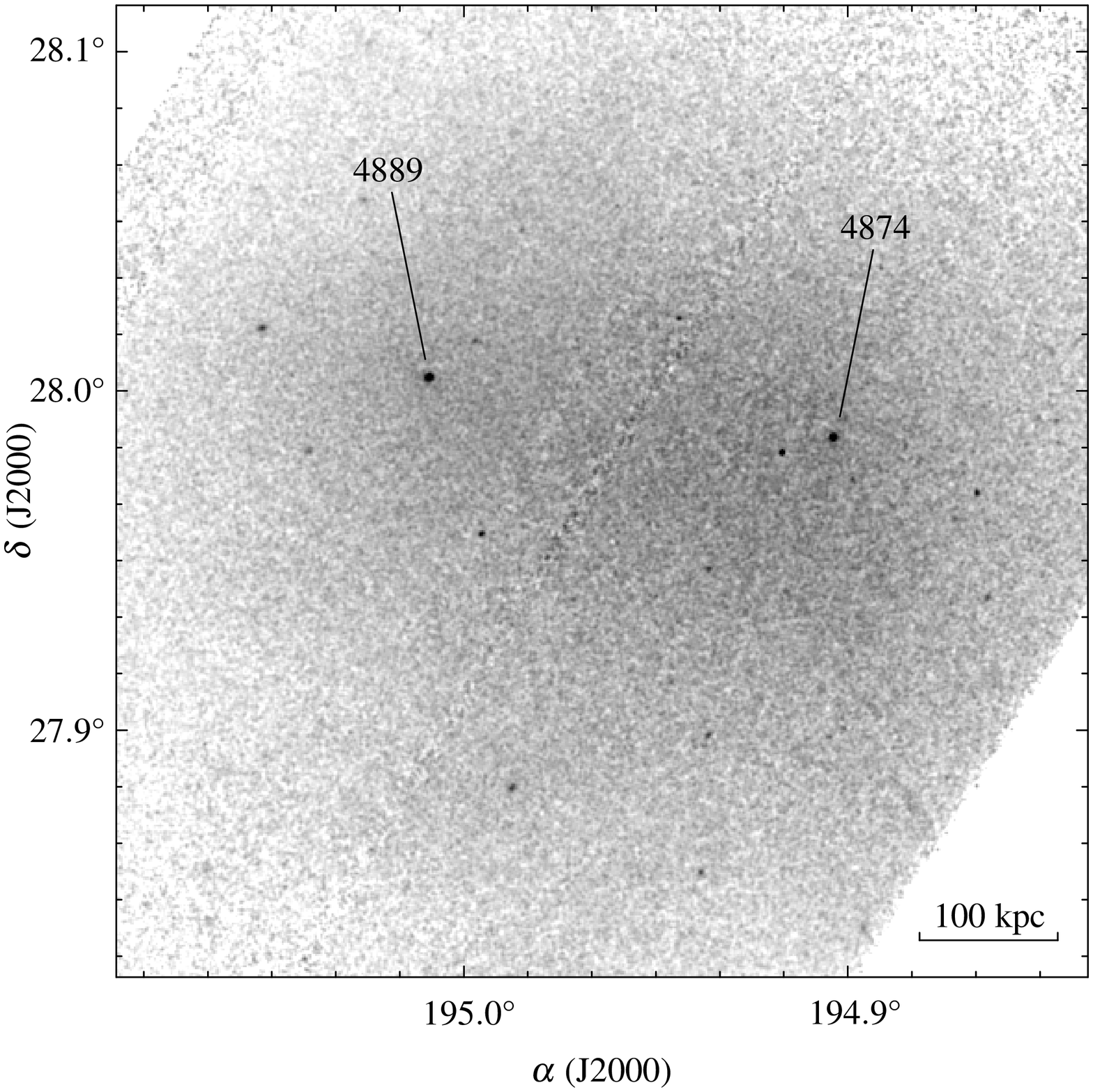}}
  \caption{The central region of the composite exposure-corrected ACIS-I
    image of Coma in the 0.5--2~keV energy band. The narrow band running
    from bottom-left to top-right is the chip gap; it disappears in the
    smoothed image. }\label{fig:big}
\end{inlinefigure}

%

The best-fit temperatures and metal abundances of these sources are listed
in Table~\ref{tab:spectra} (the uncertainties hereafter are at the 68\%
confidence level for one interesting parameter). The average temperature is
1.0~keV for \NGC4874 and 1.8~keV for \NGC4889. The average abundance of
heavy elements is close to the Solar value in both galaxies. This is
significantly higher than the metal abundance in the surrounding Coma gas
($0.25\pm0.02$, Arnaud et al.\ 2001).

\emph{Chandra}'s angular resolution permits spectral fitting in several
concentric annuli in \NGC4874. This analysis reveals a marginally
significant (at the 99\% confidence level) decrease of the gas temperature
towards the center of the galaxy (Fig.~\ref{fig:Tprofile:4874} and
Table~\ref{tab:spectra}). \NGC4889 does not show any non-isothermality, but
the statistical errors are much larger.

\begin{inlinetable}
\caption{Spectral results}\label{tab:spectra}
\def\arraystretch{1.5}
\begin{tabular}{p{3.5cm}llc}
\hline\hline
\multicolumn{1}{c}{Object} & \multicolumn{1}{c}{$kT$ (keV)} & 
\multicolumn{1}{c}{$a$ (Solar)} & $\chi^2/{\rm d.o.f}$ \\
\hline
\NGC4874\dotfill & $1.00\pm0.04$ & $0.79^{+0.84}_{-0.22}$ & 23.5/16 \\ 
\NGC4874 (0\arcsec--1.5\arcsec)\dotfill & $0.85\pm0.06$ & $0.79\,^*$ & 16.5/17\\ 
\NGC4874 (1.5\arcsec--4\arcsec)\dotfill & $1.11\pm0.08$ & $0.79\,^*$ & 14.1/17\\ 
\NGC4889 \dotfill & $1.82^{+0.22}_{-0.10}$ & $1.3^{+\infty}_{-0.4}$ &
                                            $17.9/16$ \\ 
\hline
\end{tabular}

\medskip
\begin{minipage}{0.99\linewidth}
  $^*$ --- fixed.

  The errorbars are at the $68\%$ confidence level for one interesting
  parameter. The $90\%$ lower limit on the average metal abundance is $0.43$
  and $0.81$ for \NGC4874 and \NGC4889, respectively.
\end{minipage}
\end{inlinetable}

We also checked for a possible fall in the projected temperature of the main
cluster emission on an arcmin scale around the galaxies, as observed by XMM
around \NGC4874 (Arnaud et al.\ 2001). We find no indication for cool gas
beyond the small radius of the galactic sources. In the 10\arcsec--1\arcmin\
annuli centered on the galaxies, we measure $T=8.9\pm0.4$ and
$9.2\pm0.4$~keV for \NGC4874 and 4889, respectively; these temperatures are
consistent with \emph{Chandra} and \emph{XMM} temperatures farther away, in
the 1\arcmin--3\arcmin\ annuli around the galaxies.

\section{Spatial distribution of the interstellar gas}

\begin{figure*}
  \mbox{}\hfill%
  \begin{minipage}[t]{0.45\linewidth}
    \centerline{\includegraphics[width=0.99\linewidth]{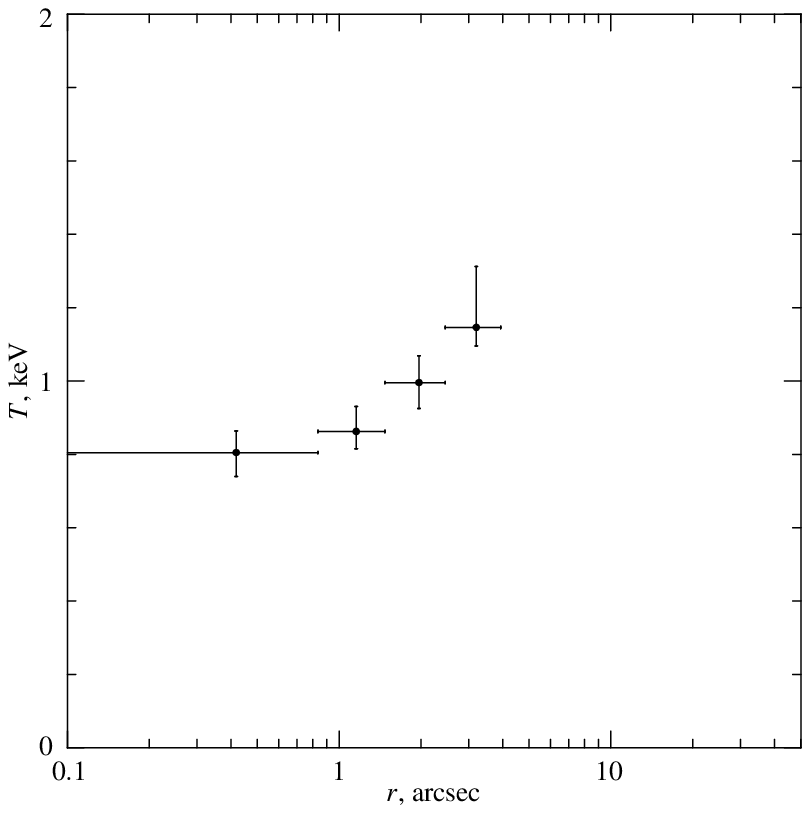}}
    \caption{Temperature profile of \NGC4874. Metal abundance and absorption
      were fixed at their average values. The statistical significance of
      the temperature difference in the radial ranges 0\arcsec--1.5\arcsec\
      and 1.5\arcsec--4\arcsec\ is 99\%. The temperature gradient can be
      modeled as $T/{\rm keV}=0.87+0.5\lg(r/{\rm arcsec})$}
    \label{fig:Tprofile:4874}
  \end{minipage}
  \hfill%
  \begin{minipage}[t]{0.45\linewidth}
    \centerline{\includegraphics[width=0.99\linewidth]{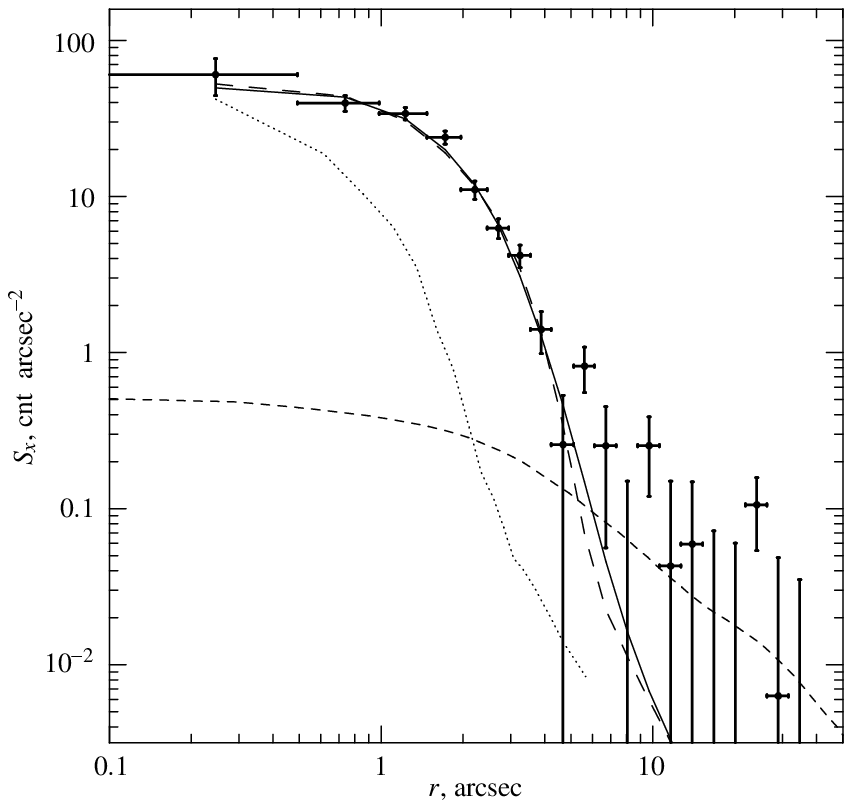}}
    \caption{X-ray surface brightness profile of \NGC4874 in the 0.5--2~keV
      band. Dotted line shows the expected radial profile for a point-like
      source; it is consistent with the observed profile of the source
      located $\approx 1'$ to the East of \NGC4874 (Fig.~\ref{fig:panels}).
      The solid line shows a $\beta$-model fit. The long-dashed line shows a
      $\beta$-model fit truncated at $r=4''$. The fits are almost
      indistinguishable because of the PSF smearing. The dashed line shows
      the HST optical surface brightness profile multiplied by the observed
      ratio $L_{\rm LMXB}/L_{\rm opt}$ (Sarazin et al.\ 2000) and therefore
      should represent the LMXB contribution to the observed X-ray flux.
      }\label{fig:profile:4874}
  \end{minipage}
  \hfill\mbox{}
\end{figure*}

The X-ray surface brightness profile of \NGC4874 is shown in
Fig.~\ref{fig:profile:4874}. A comparison with the point source profile
(dotted line) clearly shows that the observed X-ray emission is extended.
We fit the observed profiles with the $\beta$-model, $n_e = n_0
(1+r^2/r_c^2)^{-3\beta/2}$, truncated at some radius, $r_{\rm cut}$, which
is also a free parameter. The truncation of the gas distribution at $r_{\rm
cut}$ is required because the ISM is embedded in the high-pressure cluster
gas and, obviously, cannot extend beyond the radius at which the internal
and external pressures are equal. The distribution of volume emissivity was
computed assuming a constant temperature for \NGC4889 and the observed
temperature gradient for \NGC4874 (Fig.~\ref{fig:Tprofile:4874}).

We had to account for the \emph{Chandra} PSF in fitting the surface
brightness profiles because the source size is comparable to that of the
PSF. The PSF model was obtained from the library of raytracing simulations;
the PSF model is consistent with the image of the point source located
$\approx 1'$ East of \NGC4874.

\begin{inlinetable}
\caption{Gas density distribution parameters}\label{tab:profiles}
\def\arraystretch{1.5}
\begin{tabular}{p{2.4cm}llllcc}
\hline\hline
\multicolumn{1}{c}{Object} & \multicolumn{1}{c}{$n_{e0}$, cm$^{-3}$} & 
\multicolumn{1}{c}{$r_c$, kpc} & \multicolumn{1}{c}{$\beta$} &
\multicolumn{1}{c}{$r_{\rm cut}$, kpc} \\
\hline
\NGC4874\dotfill & $0.18\pm0.02$ & $0.8\pm0.3$ & $0.5\,^*$ &
$2.7\pm0.4$ \\
\NGC4874, no cut\dotfill & $0.14\pm0.02$ & $2.4\pm0.7$ & $1.8^{+\infty}_{-0.5}$ & \nodata \\ 
\NGC4889\dotfill & $0.09\pm0.01$ & $2.0\pm0.7$ &  $0.5\,^*$ & $3.1\pm0.3$
\\ 
\NGC4889, no cut\dotfill & $0.10\pm0.01$ & $3.3\pm0.7$ &  $1.7^{+\infty}_{-0.5}$ & \nodata \\ 
\hline
\end{tabular}
\medskip
\begin{minipage}{0.99\linewidth}
  $^*$ --- fixed.
\end{minipage}
\end{inlinetable}

The results of the surface brightness profile modeling are summarized in
Table~\ref{tab:profiles}. We were unable to obtain meaningful constraints on
$\beta$ because of the parameter correlation and finite PSF. Generally, all
$\beta>0.4$ are allowed. The cutoff and core-radii are well-constrained for
$\beta<1$.  For larger $\beta$, the cutoff radius becomes much larger than
the core-radius and hence has almost no effect on the observed profile. For
a pure $\beta$-model without a cutoff, $\beta$ must be in excess of
$1.2-1.3$, mimicking the cutoff at $1r_c-2r_c$, which is consistent with the
best-fit $r_{\rm cut}$. Of course, different models correspond to very similar
absolute gas density distributions in the radial range of interest.


\section{Discussion}

\subsection{Pressure Equlibrium of ISM and Cluster Gas}

It is interesting to compare the pressure of the ISM at the best-fit cutoff
radius with the external pressure of the cluster gas. Even though the
three-dimensional location of the galaxies within the cluster is uncertain,
the ICM density in their vicinity can be determined rather accurately
because each galaxy has an associated arcmin-scale X-ray enhancement
(Vikhlinin et al.\ 1994): $n_{\rm icm}=2.9\times10^{-3}$~cm$^{-3}$ around
\NGC4874 and $5.7\times10^{-3}$~cm$^{-3}$ around \NGC4889. The temperature
of the cluster gas around both galaxies is $T_{\rm icm}=9$~keV (see above).
Remarkably, we find that for the observed ISM parameters, the pressure ratio
at $r=r_{\rm cut}$ is $p_{\rm ism}/p_{\rm icm} = 1.1\pm0.3$ for \NGC4874 and
$1.2\pm0.3$ for \NGC4889.  Therefore, the pressure of the ISM in both
galaxies equals the external pressure at a radius where the \emph{Chandra}
image independently indicates an ISM density cutoff. This is a clear
indication that the distribution of the ISM is affected by the external
pressure. The derived central ISM densities are higher by a factor of 2--10
than those normally found in isolated elliptical galaxies
($n_0=0.01-0.06$~cm$^{-3}$; Forman, Jones \& Tucker 1985), which also
suggests compression by an external pressure.

\subsection{Origin and Lifetime Scales for the ISM}

Stellar ejecta is the most likely source of interstellar gas since the metal
abundance in the ISM of both galaxies is significantly higher than that in
the Coma ICM. The observed gas mass ($1.1\pm0.1\times10^8\;M_\odot$ and
$1.6\pm0.3\times10^8\;M_\odot$ in \NGC4874 and \NGC4889, respectively) would
be ejected by stars over the time $t_* = M_{\rm gas}/(\dot{M}_* L_{\rm
opt})$, where $\dot{M}_*$ is the stellar mass loss rate per unit optical
luminosity.  Using the generally adapted value $\dot{M}_* =
1.5\times10^{-11} M_\odot$~yr$^{-1}$~$L_\odot^{-1}$ (Faber \& Gallagher
1976) and the optical luminosity within $r_{\rm cut}$ (Faber et al.\ 1997),
we find $t_* \approx 8\times10^8$~yr in both galaxies. Therefore, stellar
ejecta easily produce the observed ISM over the galaxy lifetime.  The ISM is
subject to radiative cooling, thermal evaporation due to the surrounding hot
gas, heating by stellar winds and supernovae, and possibly to ram pressure
stripping.

\emph{Ram pressure stripping}. The observed X-ray centroids are within
0.5~kpc of the optical centers (limited by the \emph{Chandra} aspect
accuracy). This and the azimuthally symmetric X-ray morphology probably
exclude any significant ongoing stripping. Stripping will be unimportant if
the galaxies move together with their surrounding 100~kpc scale gas density
enhancements.

%
%

\emph{Radiative cooling}. The average cooling time of the ISM is $t_{\rm
cool} = 3 M_{\rm gas} kT /2 \mu \mp L_x$.  For the observed ISM parameters,
we find $t_{\rm cool}=1.1\times10^8$~yr for \NGC4874 and $3.7\times10^8$~yr
for \NGC4889, with a $10-20\%$ uncertainty.

\emph{Stellar winds and supernovae}. After $10^9$~years past the burst of
star formation, the average temperature of ejecta from normal stars and
supernovae is $\sim 0.6$~keV (David, Forman, \& Jones 1990). Therefore,
stellar mass loss cannot be a source of heating for the ISM with
$T=1-1.8$~keV.

\emph{Thermal evaporation}. Following Cowie \& McKee (1977), one can show
that classical heat conduction through the ISM-ICM boundaries of our
galaxies should be saturated.
From their equation (64), we find the saturated evaporation timescales
$t_{\rm evap} \approx 3\times10^6$~yr for \NGC4874 and $\approx 2\times10^6$
for \NGC4889, with a 20\% formal statistical uncertainty.

%
%
%

The timescales for radiative cooling and classical evaporation are both very
short and the ISM would be destroyed quickly by either mechanism, unless the
radiative cooling and the heat influx that drives evaporation balance each
other. This would be achieved if the conductivity through the ICM-ISM
interface is a factor of $t_{\rm cool}/t_{\rm evap}\approx 30-100$ below the
classical value. A plausible reason for such a suppression is a disjoint
magnetic structure of the ISM and ICM.
%



\subsection{Thermal Balance in \NGC4874}


If heat influx from the ICM does balance radiative cooling of the ISM, one
expects that at\pagebreak \vspace*{-55pt} each radius, the X-ray volume emissivity of the ISM equals
the divergence of the heat flux (i.e., the local
heating rate).  Interestingly, this can be verified for \NGC4874, where we
observe a temperature gradient and therefore can estimate the heat
flux. Assuming that conductivity in the ISM is classical, the heat flux is
$q=-6\times10^{-7} T^{5/2}\,dT/dr$~erg~cm$^{-2}$~s$^{-1}$ where $T$ is the
plasma temperature in degrees (McKee \& Cowie 1977). For the observed radial
dependence of temperature in \NGC4874 which can be modeled as $T(r) =
1.0\times10^7\left[1 + 0.7\lg(r/0.68\,\mbox{kpc})\right]$
(Fig.~\ref{fig:Tprofile:4874}), the divergence of $q$ is indeed within a
factor of 1.5 of the observed X-ray volume emissivity.


For \NGC4889, the uncertainty in the temperature profile is too large to
compute the heat flux. However, we can say that its ISM temperature is too
high to be plausibly maintained by internal heat sources such as stellar
mass loss and supernovae, so there should be external heating, most likely
due to conduction through the ISM boundary.

\section{Summary}

%

\emph{Chandra} observations show that the dominant Coma galaxies retain a
fraction of their X-ray coronae in the form of compact ($\sim 3$~kpc in
size) clouds with temperatures of $1.0$ and $1.8$~keV which are in the
pressure equilibrium with the hot intracluster gas. Time scales for
radiative cooling and classical evaporation are very short but the ISM
observed in two galaxies is unlikely to be a transient phenomenon.  We
propose that cooling in the ISM is balanced by heat influx through the
cluster/ISM interface. For this, thermal conductivity through the interface
should be 30--100 times below the classical value, perhaps because of the
disjoint magnetic field structure. Furthermore, in NGC4874, the observed
radial temperature gradient implies that if the conductivity {\em inside}\/
the ISM cloud is classical, then heat from outside the cloud is distributed
through the cloud volume in such a way that it precisely balances radiative
cooling at each radius. This leaves little or no energy for the ISM
evaporation, preserving the ISM clouds for a sufficiently long time.



\acknowledgments

We thank E.~Churazov for useful discussions. The financial support for this
work was provided by NASA grant NAG5-9217 and contract NAS8-39073.


\end{document}